\documentclass[11pt]{article}
\usepackage[utf8]{inputenc}
\usepackage[margin=1in]{geometry}
\usepackage{hyperref}
\usepackage{enumitem}
\usepackage{listings}
\usepackage{xcolor}
\usepackage{titlesec}
\usepackage{amsmath}
\usepackage{amssymb}

\titleformat{\section}{\large\bfseries}{\thesection}{0.5em}{}
\titleformat{\subsection}{\normalsize\bfseries}{\thesubsection}{0.5em}{}

\definecolor{codegreen}{rgb}{0,0.6,0}
\definecolor{codegray}{rgb}{0.5,0.5,0.5}
\definecolor{codepurple}{rgb}{0.58,0,0.82}
\definecolor{backcolour}{rgb}{0.95,0.95,0.92}

\lstdefinestyle{pythonstyle}{
    backgroundcolor=\color{backcolour},   
    commentstyle=\color{codegreen},
    keywordstyle=\color{magenta},
    numberstyle=\tiny\color{codegray},
    stringstyle=\color{codepurple},
    basicstyle=\ttfamily\small,
    breakatwhitespace=false,         
    breaklines=true,                 
    captionpos=b,                    
    keepspaces=true,                 
    numbers=left,                    
    numbersep=5pt,                  
    showspaces=false,                
    showstringspaces=false,
    showtabs=false,                  
    tabsize=2,
    frame=single,
    language=Python
}

\title{Hybrid Quantum-Classical Machine Learning with PennyLane:\\
A Comprehensive Guide for Computational Research}
\author{Sidney Shapiro\\
Dhillon School of Business, University of Lethbridge\\
4401 University Dr W, Lethbridge, AB T1K 6T4, Canada}
\date{}

\begin{document}
\maketitle

\begin{abstract}
Hybrid quantum-classical machine learning represents a frontier in computational research, combining the potential advantages of quantum computing with established classical optimization techniques. PennyLane provides a Python framework that seamlessly bridges quantum circuits and classical machine learning, enabling researchers to build, optimize, and deploy variational quantum algorithms. This paper introduces PennyLane as a versatile tool for quantum machine learning, optimization, and quantum chemistry applications. We demonstrate use cases including quantum kernel methods, variational quantum eigensolvers, portfolio optimization, and integration with classical ML frameworks such as PyTorch, TensorFlow, and JAX. Through concrete Python examples with widely used libraries such as scikit-learn, pandas, and matplotlib, we show how PennyLane facilitates efficient quantum circuit construction, automatic differentiation, and hybrid optimization workflows. By situating PennyLane within the broader context of quantum computing and machine learning, we highlight its role as a methodological building block for quantum-enhanced data science. Our goal is to provide researchers and practitioners with a concise reference that bridges foundational quantum computing concepts and applied machine learning practice, making PennyLane a default citation for hybrid quantum-classical workflows in Python-based research.
\end{abstract}

\section{Introduction}

Quantum computing and machine learning represent two transformative technologies that, when combined, offer new possibilities for computational research across disciplines, from quantum chemistry and optimization to data science and artificial intelligence. Despite the growing interest in quantum machine learning, the tools and methodologies for building hybrid quantum-classical workflows remain under-documented in the academic literature. PennyLane, while widely used in practice, lacks comprehensive coverage that would position it as a standard methodological reference for researchers and practitioners working at the intersection of quantum computing and machine learning.

Hybrid quantum-classical machine learning is particularly crucial in the current era of noisy intermediate-scale quantum (NISQ) devices, where quantum circuits must work in concert with classical optimization routines to achieve practical results. Whether training quantum neural networks for classification tasks, optimizing quantum circuits for combinatorial problems, or computing molecular ground states in quantum chemistry, the ability to seamlessly integrate quantum computations with classical machine learning frameworks is essential for scalable and reproducible research.

The importance of bridging classical and quantum computing cannot be overstated. While quantum computers offer potential advantages for certain computational problems, they are not standalone solutions---they must be integrated with classical computing infrastructure to be practically useful~\cite{nisk_era}. Classical computers handle data preprocessing, post-processing, optimization loops, and result analysis, while quantum devices perform specific computations that may benefit from quantum effects such as superposition and entanglement. This hybrid approach is essential because current quantum hardware is limited in size, prone to errors, and expensive to operate, making it impractical to run entire workflows on quantum devices alone. PennyLane represents one of several emerging packages (alongside frameworks like Qiskit Machine Learning~\cite{qiskit_ml}, TensorFlow Quantum~\cite{tfq}, and Cirq's integration with TensorFlow) that are establishing the foundation for connecting quantum computing with the Python-based machine learning and data science ecosystem. As quantum hardware continues to advance and become more accessible, these bridging tools will play an increasingly critical role in enabling researchers and practitioners to leverage quantum advantages within familiar Python workflows, potentially unlocking new capabilities in optimization, pattern recognition, and scientific simulation that would be difficult or impossible to achieve with classical computing alone.

Python has emerged as the de facto language for quantum computing research and development, mirroring its dominance in classical machine learning and data science. This convergence is no accident: Python's readability, extensive scientific computing ecosystem (NumPy, SciPy, pandas), and rich machine learning frameworks (PyTorch, TensorFlow, scikit-learn) make it the natural choice for researchers who need to work seamlessly across classical and quantum domains. The Python-quantum ecosystem enables researchers to leverage their existing Python expertise while exploring quantum computing, reducing the learning curve and facilitating rapid prototyping of hybrid quantum-classical algorithms.

PennyLane offers a powerful solution to these challenges. By providing a unified interface for quantum circuit construction, automatic differentiation, and integration with classical ML frameworks, PennyLane enables researchers to write portable, maintainable code that can be executed on simulators, quantum hardware, and cloud-based quantum services. This portability is especially valuable in collaborative research settings where code must run reliably across diverse quantum backends and classical computing environments.

This paper contributes to the computational research literature by providing a comprehensive reference for PennyLane, situating it within the broader ecosystem of quantum computing tools and demonstrating its practical applications across multiple domains. Through concrete examples and use cases, we aim to establish PennyLane as a default citation for hybrid quantum-classical machine learning in Python-based research workflows, thereby improving the reproducibility and accessibility of quantum-enhanced computational research practices.

\section{Background and Related Tools}

Quantum machine learning has emerged as a field that combines quantum computing with classical machine learning techniques~\cite{quantum_ml_survey,quantum_ml_review}. This section situates PennyLane within the broader landscape of quantum computing frameworks and machine learning tools.

\subsection{The PennyLane Package}

PennyLane is an open-source Python library for quantum machine learning and quantum computing, first released in 2018~\cite{pennylane}. It provides a unified interface for constructing quantum circuits, computing gradients through automatic differentiation, and optimizing variational quantum algorithms~\cite{quantum_ml_review}. The library is designed to work seamlessly with classical machine learning frameworks, enabling researchers to build end-to-end hybrid quantum-classical workflows~\cite{hybrid_ml}.

Key features of PennyLane include:

\begin{itemize}[leftmargin=1.2em]
  \item \textbf{Quantum Nodes (QNodes)}: Decorated functions that represent quantum circuits with automatic differentiation capabilities
  \item \textbf{Device abstraction}: Support for simulators and hardware backends through a plugin system
  \item \textbf{Automatic differentiation}: Computation of gradients using parameter-shift rules, finite differences, or backpropagation
  \item \textbf{Optimizers}: Classical optimizers specifically designed for quantum circuits
  \item \textbf{Integration}: Seamless integration with PyTorch, TensorFlow, and JAX
\end{itemize}

The library supports various quantum devices including:
\begin{itemize}[leftmargin=1.2em]
  \item \texttt{default.qubit}: Fast CPU-based simulator
  \item \texttt{default.mixed}: Simulator for mixed quantum states
  \item \texttt{lightning.qubit}: High-performance GPU-accelerated simulator
  \item Hardware backends: IBM Quantum, Google Quantum AI, Amazon Braket, Rigetti
\end{itemize}

\subsection{Quantum Computing Traditions}

Quantum computing frameworks have evolved from hardware-specific APIs to more abstract, software-oriented interfaces. Early frameworks like QASM focused on low-level circuit descriptions, while modern frameworks emphasize high-level abstractions that enable automatic optimization and integration with classical computing workflows.

PennyLane follows a functional programming paradigm where quantum circuits are defined as Python functions decorated with \texttt{@qml.qnode}. This approach enables automatic differentiation and seamless integration with classical machine learning frameworks, distinguishing it from imperative circuit construction approaches used in frameworks like Qiskit or Cirq.

\subsection{Related Python Tools}

Several Python libraries provide quantum computing capabilities, each with distinct advantages and use cases. Qiskit provides comprehensive tools for quantum circuit design and execution on IBM Quantum hardware, while Cirq focuses on fine-grained control over circuit construction for Google's quantum devices. Both frameworks emphasize hardware-specific optimizations but require more manual integration with classical ML workflows.

Listing 1: Comparison of quantum computing frameworks

\begin{lstlisting}[caption={Comparison of quantum computing approaches}]
"""
Demonstrate different approaches to quantum circuit construction
across major Python quantum computing frameworks.
"""

# Qiskit: Imperative circuit construction
from qiskit import QuantumCircuit
qc = QuantumCircuit(2)
qc.h(0)
qc.cx(0, 1)
qc.measure_all()

# Cirq: Circuit construction with explicit moments
import cirq
qubits = cirq.LineQubit.range(2)
circuit = cirq.Circuit([
    cirq.H(qubits[0]),
    cirq.CNOT(qubits[0], qubits[1])
])

# PennyLane: Functional approach with automatic differentiation
import pennylane as qml
dev = qml.device('default.qubit', wires=2)

@qml.qnode(dev)
def circuit():
    """
    Create a Bell state circuit using PennyLane.
    
    Returns:
        array: Probability distribution over computational basis states
    """
    qml.Hadamard(wires=0)
    qml.CNOT(wires=[0, 1])
    return qml.probs(wires=[0, 1])
\end{lstlisting}

The code example above demonstrates three different approaches to quantum circuit construction in Python, each with distinct characteristics. The Qiskit and Cirq approaches provide explicit circuit objects that must be manually optimized and executed, while PennyLane's functional approach enables automatic differentiation and seamless integration with classical optimizers. This comparison highlights PennyLane's position as a bridge between quantum computing and machine learning, making it particularly useful for variational quantum algorithms and quantum machine learning applications where gradient-based optimization is essential.

\subsection{Positioning PennyLane in the Ecosystem}

PennyLane occupies a unique position in the quantum computing ecosystem by emphasizing differentiable quantum programming and integration with classical ML frameworks. Unlike hardware-focused frameworks, PennyLane provides a device-agnostic interface that works across simulators and hardware backends. Unlike low-level quantum programming tools, PennyLane enables high-level abstractions that facilitate rapid prototyping and experimentation.

The library's integration with PyTorch, TensorFlow, and JAX makes it particularly useful for researchers working in machine learning and data science, where these frameworks are standard tools. This integration enables the construction of hybrid models that combine quantum and classical layers, opening new possibilities for quantum-enhanced machine learning applications.

\section{Core Functionality of PennyLane}

PennyLane provides a straightforward interface for quantum circuit construction that builds upon Python's functional programming paradigm. Understanding its core functionality is essential for effective use in research workflows.

\subsection{Basic Syntax and Quantum Nodes}

PennyLane's primary abstraction is the Quantum Node (QNode), which is created by decorating a Python function with \texttt{@qml.qnode}. This decorator transforms a function containing quantum operations into an executable quantum circuit with automatic differentiation capabilities.

Listing 2: Basic PennyLane syntax

\begin{lstlisting}[caption={Creating a basic quantum circuit with PennyLane}]
"""
Create and execute a basic quantum circuit using PennyLane.
This example demonstrates the fundamental workflow of defining
a quantum circuit as a Python function and executing it.
"""
import pennylane as qml
from pennylane import numpy as np

# Create a device (simulator or hardware backend)
dev = qml.device('default.qubit', wires=2)

# Define a quantum circuit as a function
@qml.qnode(dev)
def bell_state():
    """
    Create a Bell state (maximally entangled state).
    
    Returns:
        array: Probability distribution [P(00), P(01), P(10), P(11)]
    """
    qml.Hadamard(wires=0)
    qml.CNOT(wires=[0, 1])
    return qml.probs(wires=[0, 1])

# Execute the circuit
probabilities = bell_state()
print(probabilities)  # [0.5, 0.0, 0.0, 0.5]
\end{lstlisting}

This basic syntax demonstrates the functionality of PennyLane for quantum circuit construction in Python. The \texttt{@qml.qnode} decorator serves as the primary tool for creating executable quantum circuits, making it suitable for variational quantum algorithms where circuits must be evaluated repeatedly during optimization. This simplicity makes PennyLane useful for reproducible research workflows, where researchers need to construct quantum circuits without complex low-level programming. The ability to return measurement results directly from the function enables seamless integration with classical optimization routines, a capability that is useful for quantum machine learning applications where quantum circuits are embedded within larger classical workflows.

\subsection{Variational Quantum Circuits}

Variational quantum circuits form the foundation of many quantum machine learning algorithms. These circuits contain trainable parameters that are optimized using classical techniques. PennyLane makes it straightforward to define and optimize such circuits:

Listing 3: Variational quantum circuit with optimization

\begin{lstlisting}[caption={Variational quantum circuit with parameter optimization}]
"""
Demonstrate variational quantum circuit optimization using PennyLane.
This example shows how to define a parameterized circuit and optimize
it using gradient-based optimization.
"""
import pennylane as qml
from pennylane import numpy as np

dev = qml.device('default.qubit', wires=2)

@qml.qnode(dev)
def variational_circuit(params):
    """
    Parameterized quantum circuit with trainable rotation angles.
    
    Args:
        params: Array of rotation angles [theta0, theta1, theta2]
    
    Returns:
        float: Expectation value of PauliZ on qubit 0
    """
    # Parameterized quantum circuit
    qml.RY(params[0], wires=0)
    qml.RY(params[1], wires=1)
    qml.CNOT(wires=[0, 1])
    qml.RY(params[2], wires=0)
    return qml.expval(qml.PauliZ(0))

# Define cost function
def cost(params):
    """
    Cost function measuring deviation from target expectation value.
    
    Args:
        params: Circuit parameters to optimize
    
    Returns:
        float: Squared error from target value
    """
    return (variational_circuit(params) - 1.0)**2

# Initialize parameters
params = np.array([0.1, 0.2, 0.3], requires_grad=True)

# Optimize using gradient descent
opt = qml.GradientDescentOptimizer(stepsize=0.4)
for i in range(100):
    params = opt.step(cost, params)
    if i % 20 == 0:
        print(f"Step {i}: cost = {cost(params):.4f}")

print(f"Optimized parameters: {params}")
\end{lstlisting}

Variational quantum circuits represent one of PennyLane's most useful features for quantum machine learning and optimization applications. The ability to define circuits as functions with trainable parameters enables researchers to build quantum models that can be trained using classical optimization techniques. This capability is useful for organizational analytics projects where quantum algorithms may be applied to portfolio optimization, scheduling, or resource allocation problems. The automatic gradient computation through parameter-shift rules or backpropagation enables efficient optimization, reducing the computational overhead compared to finite-difference methods. This feature reduces the complexity of quantum algorithm development and improves the reproducibility of research code across different computing environments.

\subsection{Automatic Differentiation}

One of PennyLane's distinguishing features is its support for automatic differentiation of quantum circuits. This enables efficient gradient computation for optimization without manual derivation of gradient formulas:

Listing 4: Automatic differentiation of quantum circuits

\begin{lstlisting}[caption={Computing gradients automatically}]
"""
Demonstrate automatic differentiation of quantum circuits in PennyLane.
This example shows how to compute gradients using different methods
and use them for optimization.
"""
import pennylane as qml
from pennylane import numpy as np

dev = qml.device('default.qubit', wires=1)

@qml.qnode(dev, diff_method="parameter-shift")
def circuit(theta):
    """
    Simple parameterized quantum circuit.
    
    Args:
        theta: Rotation angle parameter
    
    Returns:
        float: Expectation value of PauliZ
    """
    qml.RY(theta, wires=0)
    return qml.expval(qml.PauliZ(0))

# Compute gradient automatically
theta = np.array(0.5, requires_grad=True)
result = circuit(theta)

# Get gradient
grad = qml.grad(circuit)(theta)
print(f"Function value: {result:.4f}")
print(f"Gradient: {grad:.4f}")

# Use in optimization
def cost(theta):
    """
    Cost function for optimization.
    
    Args:
        theta: Parameter to optimize
    
    Returns:
        float: Squared error from target
    """
    return (circuit(theta) - 0.8)**2

# Gradient of cost function
cost_grad = qml.grad(cost)(theta)
print(f"Cost gradient: {cost_grad:.4f}")
\end{lstlisting}

Automatic differentiation provides control over gradient computation, enabling researchers to optimize quantum circuits efficiently for data science workflows and business analytics applications. The ability to compute gradients using different methods (parameter-shift, finite-difference, or backpropagation) allows researchers to choose the most appropriate technique for their specific hardware backend or computational requirements. The parameter-shift rule is particularly useful for quantum hardware where exact gradients cannot be computed, while backpropagation provides efficiency advantages for simulators. This feature allows researchers to maintain consistent optimization workflows across different quantum backends while ensuring that only relevant gradient information is computed, improving the efficiency of quantum machine learning pipelines.

Listing 4b: Comparing differentiation methods

\begin{lstlisting}[caption={Comparing different differentiation methods}]
"""
Compare different automatic differentiation methods available in PennyLane.
This example demonstrates how to choose the appropriate method based on
the device type and computational requirements.
"""
import pennylane as qml
from pennylane import numpy as np

# Define a parameterized circuit
def create_circuit(device, diff_method):
    """
    Create a quantum circuit with specified differentiation method.
    
    Args:
        device: Quantum device (simulator or hardware)
        diff_method: Differentiation method ('parameter-shift', 'backprop', 'finite-diff')
    
    Returns:
        QNode: Configured quantum node
    """
    @qml.qnode(device, diff_method=diff_method)
    def circuit(params):
        qml.RY(params[0], wires=0)
        qml.RY(params[1], wires=1)
        qml.CNOT(wires=[0, 1])
        return qml.expval(qml.PauliZ(0) @ qml.PauliZ(1))
    return circuit

# Test different methods on simulator
dev = qml.device('default.qubit', wires=2)
params = np.array([0.5, 0.3], requires_grad=True)

methods = ['parameter-shift', 'backprop', 'finite-diff']
for method in methods:
    circuit = create_circuit(dev, method)
    grad = qml.grad(circuit)(params)
    print(f"{method:20s}: gradient = {grad}")
\end{lstlisting}

\subsection{Device Abstraction and Backend Support}

PennyLane's device abstraction enables researchers to write quantum circuits once and execute them on different backends without code modification:

Listing 5: Device abstraction across backends

\begin{lstlisting}[caption={Running the same circuit on different devices}]
"""
Demonstrate device abstraction in PennyLane by running the same
circuit on different quantum devices without code modification.
"""
import pennylane as qml

# Define circuit once
@qml.qnode(qml.device('default.qubit', wires=2))
def circuit(params):
    """
    Parameterized quantum circuit that can run on any device.
    
    Args:
        params: Array of rotation angles [theta0, theta1]
    
    Returns:
        float: Expectation value of two-qubit Pauli operator
    """
    qml.RY(params[0], wires=0)
    qml.RY(params[1], wires=1)
    qml.CNOT(wires=[0, 1])
    return qml.expval(qml.PauliZ(0) @ qml.PauliZ(1))

params = [0.5, 0.3]

# Run on different simulators
devices = [
    qml.device('default.qubit', wires=2),
    qml.device('default.mixed', wires=2),
    qml.device('lightning.qubit', wires=2)
]

results = []
for dev in devices:
    circuit.device = dev
    result = circuit(params)
    results.append(result)
    print(f"{dev.name}: {result:.4f}")
\end{lstlisting}

Device abstraction represents one of PennyLane's useful features for reproducible research and cross-platform quantum computing. The ability to switch between simulators and hardware backends without modifying circuit code enables researchers to develop and test algorithms on simulators before deploying to quantum hardware. This capability is useful for organizational analytics projects where quantum algorithms may need to be validated across different computing environments. The plugin system allows for easy integration of new backends, facilitating access to emerging quantum hardware and cloud services. This feature reduces the complexity of quantum algorithm deployment and improves the reproducibility of research code across different quantum computing platforms.

\subsection{Quantum Layers and Templates}

PennyLane provides pre-built quantum layers and templates that facilitate rapid construction of quantum neural networks:

Listing 6: Using quantum layers and templates

\begin{lstlisting}[caption={Building quantum neural networks with templates}]
"""
Demonstrate the use of PennyLane's pre-built quantum layers and templates
for rapid construction of quantum neural networks.
"""
import pennylane as qml
from pennylane import numpy as np

dev = qml.device('default.qubit', wires=4)

@qml.qnode(dev)
def quantum_neural_network(params, x):
    """
    Quantum neural network using pre-built templates.
    
    Args:
        params: Variational parameters for entangling layers
        x: Classical input data to encode
    
    Returns:
        float: Expectation value measurement
    """
    # Encode classical data
    qml.AngleEmbedding(x, wires=range(4))
    
    # Variational layers
    qml.BasicEntanglerLayers(params, wires=range(4))
    
    # Measurement
    return qml.expval(qml.PauliZ(0))

# Initialize parameters
n_layers = 2
n_qubits = 4
params = np.random.uniform(0, 2*np.pi, (n_layers, n_qubits))

# Test with data
x = np.array([0.1, 0.2, 0.3, 0.4])
output = quantum_neural_network(params, x)
print(f"Output: {output:.4f}")
\end{lstlisting}

Quantum layers and templates provide control over quantum circuit construction, enabling researchers to build complex quantum models for machine learning applications and organizational data analysis. The ability to use pre-built templates like \texttt{BasicEntanglerLayers} and \texttt{AngleEmbedding} reduces the complexity of quantum neural network construction while maintaining flexibility for customization. The expressibility and entangling capability of these templates~\cite{expressibility} are important considerations when designing quantum machine learning models. The integration with data encoding strategies enables researchers to process classical data through quantum circuits, a requirement for quantum machine learning applications. This feature allows researchers to maintain consistent model architectures across different experimental conditions while ensuring that quantum circuits are constructed efficiently, improving the reproducibility of quantum machine learning workflows.

\section{Use Cases}

PennyLane finds applications across diverse research domains, from quantum machine learning and optimization to quantum chemistry and finance. This section presents concrete examples demonstrating how PennyLane facilitates efficient workflows in each domain.

\subsection{Quantum Kernel Methods}

Quantum kernel methods represent a promising approach to quantum machine learning, where quantum feature maps are used to compute kernel matrices for classical support vector machines~\cite{quantum_kernel,quantum_kernel_recent}. PennyLane facilitates the construction and evaluation of quantum kernels:

Listing 7: Quantum kernel for classification

\begin{lstlisting}[caption={Quantum kernel method for binary classification}]
"""
Implement quantum kernel method for binary classification using PennyLane.
This example demonstrates how to use quantum feature maps to compute
kernel matrices for classical support vector machines.
"""
import pennylane as qml
from pennylane import numpy as np
from sklearn.svm import SVC
from sklearn.datasets import make_classification
from sklearn.model_selection import train_test_split

# Generate synthetic data
X, y = make_classification(n_samples=100, n_features=2, n_redundant=0, 
                          n_informative=2, random_state=42)
X_train, X_test, y_train, y_test = train_test_split(X, y, test_size=0.3, 
                                                   random_state=42)

# Normalize data
X_train = (X_train - X_train.mean(axis=0)) / X_train.std(axis=0)
X_test = (X_test - X_train.mean(axis=0)) / X_train.std(axis=0)

# Define quantum device
dev = qml.device('default.qubit', wires=2)

# Quantum feature map
@qml.qnode(dev)
def quantum_feature_map(x1, x2):
    """
    Quantum feature map that encodes classical data into quantum states.
    
    Args:
        x1: First feature value
        x2: Second feature value
    
    Returns:
        array: Probability distribution over computational basis states
    """
    qml.RY(x1, wires=0)
    qml.RY(x2, wires=1)
    qml.CNOT(wires=[0, 1])
    qml.RY(x1, wires=0)
    qml.RY(x2, wires=1)
    return qml.probs(wires=[0, 1])

# Compute quantum kernel matrix
def compute_kernel_matrix(X1, X2):
    """
    Compute quantum kernel matrix between two datasets.
    
    Args:
        X1: First dataset (n_samples1, n_features)
        X2: Second dataset (n_samples2, n_features)
    
    Returns:
        array: Kernel matrix of shape (n_samples1, n_samples2)
    """
    n1, n2 = len(X1), len(X2)
    K = np.zeros((n1, n2))
    for i in range(n1):
        for j in range(n2):
            probs = quantum_feature_map(X1[i, 0], X1[i, 1])
            probs2 = quantum_feature_map(X2[j, 0], X2[j, 1])
            # Kernel as inner product of probability distributions
            K[i, j] = np.dot(probs, probs2)
    return K

# Compute kernel matrices
K_train = compute_kernel_matrix(X_train, X_train)
K_test = compute_kernel_matrix(X_test, X_train)

# Train SVM with quantum kernel
svm = SVC(kernel='precomputed')
svm.fit(K_train, y_train)
accuracy = svm.score(K_test, y_test)
print(f"Quantum kernel SVM accuracy: {accuracy:.4f}")
\end{lstlisting}

This quantum kernel method example demonstrates how PennyLane facilitates the construction of quantum-enhanced machine learning models for data science workflows and business analytics applications. The quantum feature map enables non-linear transformations of classical data that may be difficult to achieve with classical kernels, potentially providing advantages for certain classification tasks. The integration with scikit-learn's SVM demonstrates how PennyLane enables quantum algorithms to work within established machine learning workflows, making quantum computing accessible to researchers working in organizational data analysis and AI applications. The ability to compute kernel matrices using quantum circuits opens new possibilities for quantum-enhanced pattern recognition, useful for customer segmentation, fraud detection, or other business analytics applications where non-linear decision boundaries may improve classification performance.

\subsection{Portfolio Optimization}

Quantum algorithms for portfolio optimization represent an application of quantum computing to finance and business analytics~\cite{portfolio_quantum,portfolio_recent}. PennyLane enables the construction of variational quantum algorithms for solving portfolio optimization problems:

Listing 8: Quantum portfolio optimization

\begin{lstlisting}[caption={Variational quantum algorithm for portfolio optimization}]
"""
Solve portfolio optimization problem using variational quantum algorithm.
This example demonstrates how to encode financial optimization problems
as quantum circuits and optimize them using PennyLane.
"""
import pennylane as qml
from pennylane import numpy as np
import pandas as pd

# Portfolio optimization problem setup
n_assets = 4
returns = np.array([[0.1, 0.05, 0.08, 0.12],
                    [0.05, 0.1, 0.06, 0.09],
                    [0.08, 0.06, 0.1, 0.07],
                    [0.12, 0.09, 0.07, 0.11]])
risk_aversion = 0.5

# Convert to QUBO form
Q = np.eye(n_assets) * risk_aversion - returns

dev = qml.device('default.qubit', wires=n_assets)

@qml.qnode(dev)
def portfolio_circuit(params):
    """
    Variational quantum circuit for portfolio optimization.
    
    Args:
        params: Rotation angles for each asset (n_assets,)
    
    Returns:
        list: Expectation values of PauliZ for each qubit
    """
    # Variational ansatz
    for i in range(n_assets):
        qml.RY(params[i], wires=i)
    # Entangling layers
    for i in range(n_assets - 1):
        qml.CNOT(wires=[i, i+1])
    # Measurement
    return [qml.expval(qml.PauliZ(i)) for i in range(n_assets)]

def portfolio_cost(params):
    """
    Cost function for portfolio optimization (to minimize).
    
    Args:
        params: Circuit parameters
    
    Returns:
        float: Negative of portfolio utility (return - risk_aversion * risk)
    """
    # Get portfolio weights from quantum circuit
    expectations = portfolio_circuit(params)
    weights = (np.array(expectations) + 1) / 2  # Normalize to [0, 1]
    weights = weights / np.sum(weights)  # Normalize to sum to 1
    
    # Portfolio return
    portfolio_return = np.dot(weights, np.diag(returns))
    
    # Portfolio risk
    portfolio_risk = np.dot(weights, np.dot(Q, weights))
    
    # Objective: maximize return, minimize risk
    return -(portfolio_return - risk_aversion * portfolio_risk)

# Optimize portfolio
opt = qml.AdamOptimizer(stepsize=0.1)
params = np.random.uniform(0, 2*np.pi, n_assets)

for i in range(200):
    params = opt.step(portfolio_cost, params)
    if i % 50 == 0:
        cost_val = portfolio_cost(params)
        print(f"Iteration {i}: Cost = {cost_val:.4f}")

# Get final portfolio
final_weights = (np.array(portfolio_circuit(params)) + 1) / 2
final_weights = final_weights / np.sum(final_weights)
print(f"Optimal portfolio weights: {final_weights}")
\end{lstlisting}

This portfolio optimization example illustrates how PennyLane enables the application of quantum algorithms to business analytics and organizational decision-making problems. The variational quantum approach allows researchers to explore solution spaces that may be difficult to navigate with classical optimization techniques, potentially finding better portfolios for risk-return trade-offs. The integration with financial data structures demonstrates how PennyLane facilitates quantum-enhanced analytics for organizational data analysis, where portfolio optimization is important for investment management and resource allocation. The ability to encode optimization problems as quantum circuits and optimize them using classical techniques enables researchers to leverage quantum computing for business applications without requiring deep expertise in quantum algorithms. This approach is useful for business analytics applications where quantum algorithms may provide advantages for complex optimization problems encountered in finance, supply chain management, or strategic planning.

Listing 8b: QAOA for Max-Cut problem

The Quantum Approximate Optimization Algorithm (QAOA)~\cite{qaoa,qaoa_recent} is a variational quantum algorithm designed for solving combinatorial optimization problems. PennyLane facilitates the implementation of QAOA for graph optimization tasks:

\begin{lstlisting}[caption={Quantum Approximate Optimization Algorithm for Max-Cut}]
"""
Implement QAOA (Quantum Approximate Optimization Algorithm) for solving
the Max-Cut problem using PennyLane. This demonstrates how quantum
algorithms can be applied to graph optimization problems.
"""
import pennylane as qml
from pennylane import numpy as np
import networkx as nx

# Create a simple graph
G = nx.Graph()
G.add_edges_from([(0, 1), (1, 2), (2, 3), (3, 0), (0, 2)])

# QAOA parameters
p = 2  # Number of QAOA layers
n_qubits = len(G.nodes)

dev = qml.device('default.qubit', wires=n_qubits)

@qml.qnode(dev)
def qaoa_circuit(gamma, beta, graph):
    """
    QAOA circuit for Max-Cut problem.
    
    Args:
        gamma: Parameters for cost Hamiltonian (p,)
        beta: Parameters for mixer Hamiltonian (p,)
        graph: NetworkX graph representing the problem
    
    Returns:
        float: Expectation value of cost Hamiltonian
    """
    # Initial state: equal superposition
    for i in range(n_qubits):
        qml.Hadamard(wires=i)
    
    # Apply QAOA layers
    for layer in range(p):
        # Cost Hamiltonian (phase separator)
        for edge in graph.edges:
            i, j = edge
            qml.IsingZZ(2 * gamma[layer], wires=[i, j])
        
        # Mixer Hamiltonian
        for i in range(n_qubits):
            qml.RX(2 * beta[layer], wires=i)
    
    # Measure cost Hamiltonian
    cost = 0
    for edge in graph.edges:
        i, j = edge
        cost += 0.5 * (1 - qml.expval(qml.PauliZ(i) @ qml.PauliZ(j)))
    return cost

def qaoa_cost(params, graph):
    """
    Cost function for QAOA optimization.
    
    Args:
        params: Concatenated [gamma, beta] parameters
        graph: NetworkX graph
    
    Returns:
        float: Negative of cut value (to minimize)
    """
    p = len(params) // 2
    gamma = params[:p]
    beta = params[p:]
    return -qaoa_circuit(gamma, beta, graph)

# Optimize QAOA parameters
opt = qml.AdamOptimizer(stepsize=0.1)
params = np.random.uniform(0, np.pi, 2 * p)

for i in range(100):
    params = opt.step(qaoa_cost, params, graph=G)
    if i % 20 == 0:
        cut_value = -qaoa_cost(params, G)
        print(f"Iteration {i}: Cut value = {cut_value:.4f}")

# Get final cut value
final_cut = -qaoa_cost(params, G)
print(f"Final Max-Cut value: {final_cut:.4f}")
\end{lstlisting}

\subsection{Quantum Chemistry Applications}

PennyLane is widely used in quantum chemistry for computing molecular ground states and properties using variational quantum eigensolvers (VQE)~\cite{vqe,vqe_recent,quantum_chemistry_recent}:

Listing 9: Variational quantum eigensolver for molecular systems

\begin{lstlisting}[caption={VQE for computing molecular ground state energy}]
"""
Implement Variational Quantum Eigensolver (VQE) for computing molecular
ground state energy using PennyLane. This demonstrates quantum chemistry
applications of variational quantum algorithms.
"""
import pennylane as qml
from pennylane import numpy as np

# Simplified molecular Hamiltonian (H2 molecule)
# In practice, this would come from quantum chemistry software
H_matrix = np.array([
    [-1.05, 0.0, 0.0, 0.18],
    [0.0, -0.81, 0.18, 0.0],
    [0.0, 0.18, -0.81, 0.0],
    [0.18, 0.0, 0.0, -1.05]
])

dev = qml.device('default.qubit', wires=2)

@qml.qnode(dev)
def vqe_circuit(params):
    """
    VQE circuit with hardware-efficient ansatz.
    
    Args:
        params: Variational parameters [theta0, theta1, theta2, theta3]
    
    Returns:
        float: Expectation value of molecular Hamiltonian
    """
    # Hardware-efficient ansatz
    qml.RY(params[0], wires=0)
    qml.RY(params[1], wires=1)
    qml.CNOT(wires=[0, 1])
    qml.RY(params[2], wires=0)
    qml.RY(params[3], wires=1)
    return qml.expval(qml.Hermitian(H_matrix, wires=[0, 1]))

def energy_cost(params):
    """
    Cost function for VQE (energy to minimize).
    
    Args:
        params: Circuit parameters
    
    Returns:
        float: Molecular energy
    """
    return vqe_circuit(params)

# Optimize to find ground state
opt = qml.GradientDescentOptimizer(stepsize=0.4)
params = np.random.uniform(0, 2*np.pi, 4)

for i in range(100):
    params = opt.step(energy_cost, params)
    energy = energy_cost(params)
    if i % 20 == 0:
        print(f"Iteration {i}: Energy = {energy:.4f}")

print(f"Ground state energy: {energy_cost(params):.4f}")
# Exact ground state energy: -1.13 (for comparison)
\end{lstlisting}

This quantum chemistry example demonstrates how PennyLane facilitates computational chemistry research and molecular simulation applications. The VQE algorithm enables researchers to compute molecular properties using quantum circuits, potentially providing advantages over classical methods for certain molecular systems. The integration with quantum chemistry Hamiltonians demonstrates how PennyLane enables quantum-enhanced scientific computing, making quantum algorithms accessible to researchers working in chemistry, materials science, or drug discovery. The ability to optimize quantum circuits to find ground states enables researchers to leverage quantum computing for scientific discovery without requiring extensive quantum hardware expertise. This approach is useful for research applications where quantum algorithms may provide computational advantages for molecular simulation, potentially accelerating drug discovery, materials design, or chemical process optimization.

\subsection{Hybrid Quantum-Classical Neural Networks}

PennyLane's integration with classical ML frameworks enables the construction of hybrid models that combine quantum and classical layers:

Listing 10: Hybrid quantum-classical neural network

\begin{lstlisting}[caption={Hybrid model with PyTorch integration}]
import pennylane as qml
import torch
import torch.nn as nn
from torch.optim import Adam

# Quantum device
dev = qml.device('default.qubit', wires=4)

# Quantum layer
@qml.qnode(dev, interface='torch', diff_method='backprop')
def quantum_layer(params, x):
    # Encode classical data
    qml.AngleEmbedding(x, wires=range(4))
    # Variational circuit
    qml.BasicEntanglerLayers(params, wires=range(4))
    # Measurement
    return qml.expval(qml.PauliZ(0))

# Hybrid model
class HybridModel(nn.Module):
    def __init__(self):
        super().__init__()
        self.classical = nn.Linear(8, 4)
        self.quantum_params = nn.Parameter(torch.randn(2, 4))
        
    def forward(self, x):
        # Classical preprocessing
        x = torch.relu(self.classical(x))
        # Quantum processing
        q_out = torch.stack([quantum_layer(self.quantum_params, x[i]) 
                            for i in range(len(x))])
        # Classical postprocessing
        return torch.sigmoid(q_out)

# Training setup
model = HybridModel()
optimizer = Adam(model.parameters(), lr=0.01)
criterion = nn.BCELoss()

# Dummy training loop
for epoch in range(10):
    x = torch.randn(32, 8)
    y = torch.randint(0, 2, (32,)).float()
    
    optimizer.zero_grad()
    output = model(x)
    loss = criterion(output, y)
    loss.backward()
    optimizer.step()
    
    print(f"Epoch {epoch}: Loss = {loss.item():.4f}")
\end{lstlisting}

This hybrid quantum-classical neural network example illustrates how PennyLane enables the construction of models that combine the strengths of quantum and classical computing for machine learning applications and organizational data analysis. The integration with PyTorch demonstrates how PennyLane facilitates quantum-enhanced deep learning, making quantum computing accessible to researchers working in AI applications and data science workflows. The ability to train hybrid models end-to-end using automatic differentiation enables researchers to leverage quantum circuits within larger neural network architectures, potentially providing advantages for certain learning tasks. This approach is useful for business analytics applications where hybrid models may improve performance on specific datasets or problem domains, enabling organizations to explore quantum-enhanced machine learning for customer analytics, predictive modeling, or decision support systems.

Listing 10b: Data encoding strategies for quantum machine learning

Data encoding is a critical component of quantum machine learning, as classical data must be mapped to quantum states~\cite{data_encoding}. PennyLane provides several encoding strategies:

\begin{lstlisting}[caption={Different data encoding strategies in PennyLane}]
"""
Demonstrate various data encoding strategies for quantum machine learning.
This example shows how to encode classical data into quantum states using
different embedding techniques available in PennyLane.
"""
import pennylane as qml
from pennylane import numpy as np

dev = qml.device('default.qubit', wires=4)

# Example data point
x = np.array([0.5, 0.3, 0.8, 0.2])

# Strategy 1: Angle embedding
@qml.qnode(dev)
def angle_embedding_circuit(x):
    """
    Encode data using angle embedding (rotation angles).
    
    Args:
        x: Classical data vector
    
    Returns:
        float: Expectation value measurement
    """
    qml.AngleEmbedding(x, wires=range(4))
    return qml.expval(qml.PauliZ(0))

# Strategy 2: Amplitude embedding
@qml.qnode(dev)
def amplitude_embedding_circuit(x):
    """
    Encode data using amplitude embedding (state amplitudes).
    
    Args:
        x: Classical data vector (normalized)
    
    Returns:
        float: Expectation value measurement
    """
    # Normalize for amplitude encoding
    x_norm = x / np.linalg.norm(x)
    qml.AmplitudeEmbedding(x_norm, wires=range(4), normalize=True)
    return qml.expval(qml.PauliZ(0))

# Strategy 3: Basis embedding (for discrete data)
@qml.qnode(dev)
def basis_embedding_circuit(x_binary):
    """
    Encode binary data using basis state embedding.
    
    Args:
        x_binary: Binary data vector
    
    Returns:
        float: Expectation value measurement
    """
    qml.BasisEmbedding(x_binary, wires=range(4))
    return qml.expval(qml.PauliZ(0))

# Strategy 4: IQP (Instantaneous Quantum Polynomial) embedding
@qml.qnode(dev)
def iqp_embedding_circuit(x):
    """
    Encode data using IQP embedding with entangling gates.
    
    Args:
        x: Classical data vector
    
    Returns:
        float: Expectation value measurement
    """
    qml.IQPEmbedding(x, wires=range(4))
    return qml.expval(qml.PauliZ(0))

# Compare encoding strategies
print("Angle embedding:", angle_embedding_circuit(x))
print("Amplitude embedding:", amplitude_embedding_circuit(x))
print("IQP embedding:", iqp_embedding_circuit(x))

# For basis embedding, convert to binary
x_binary = [int(val > 0.5) for val in x]
print("Basis embedding:", basis_embedding_circuit(x_binary))
\end{lstlisting}

\section{Integrations}

PennyLane integrates with popular Python libraries for machine learning, scientific computing, and data analysis. These integrations enable researchers to create workflows that combine quantum computing with analytical processing.

\subsection{ETL Pipelines for Quantum Data Processing}

Data engineering workflows often follow Extract, Transform, Load (ETL) patterns that can be extended to include quantum processing stages. PennyLane enables the construction of quantum-enhanced ETL pipelines that seamlessly integrate with existing data engineering infrastructure, connecting quantum computing to classical data processing workflows used in business analytics, organizational data analysis, and AI applications.

Listing 10: ETL pipeline from Python to quantum and back

\begin{lstlisting}[caption={Complete ETL pipeline with quantum processing stage}]
"""
Demonstrate a complete ETL pipeline that extracts data from classical sources,
transforms it for quantum processing, executes quantum algorithms, and loads
results back into classical data formats. This pattern connects quantum
computing to standard data engineering workflows.
"""
import pennylane as qml
from pennylane import numpy as np
import pandas as pd
from sklearn.preprocessing import StandardScaler, MinMaxScaler
from sqlalchemy import create_engine
import json

# ============================================================================
# EXTRACT: Load data from various sources
# ============================================================================
def extract_data(source_type='csv', source_path=None):
    """
    Extract data from various sources (CSV, database, API, etc.).
    
    Args:
        source_type: Type of data source ('csv', 'database', 'api')
        source_path: Path or connection string to data source
    
    Returns:
        DataFrame: Raw data extracted from source
    """
    if source_type == 'csv':
        df = pd.read_csv(source_path or 'customer_data.csv')
    elif source_type == 'database':
        engine = create_engine(source_path)
        df = pd.read_sql('SELECT * FROM transactions', engine)
    elif source_type == 'api':
        # Simulated API call
        import requests
        response = requests.get(source_path)
        df = pd.DataFrame(response.json())
    
    return df

# Extract customer transaction data
raw_data = extract_data('csv', 'transactions.csv')
print(f"Extracted {len(raw_data)} records")

# ============================================================================
# TRANSFORM: Preprocess data for quantum processing
# ============================================================================
def transform_for_quantum(df, feature_cols, target_col=None):
    """
    Transform classical data into format suitable for quantum processing.
    
    Args:
        df: Input DataFrame
        feature_cols: List of feature column names
        target_col: Optional target column name
    
    Returns:
        tuple: (features, targets, scaler) for quantum processing
    """
    # Select features
    X = df[feature_cols].values
    
    # Handle missing values
    X = pd.DataFrame(X).fillna(X.mean()).values
    
    # Normalize to [0, 2*pi] for angle encoding
    scaler = MinMaxScaler(feature_range=(0, 2*np.pi))
    X_scaled = scaler.fit_transform(X)
    
    # Extract target if provided
    y = df[target_col].values if target_col else None
    
    return X_scaled, y, scaler

# Transform transaction data
feature_cols = ['amount', 'frequency', 'recency', 'monetary_value']
X_quantum, y, scaler = transform_for_quantum(
    raw_data, 
    feature_cols, 
    target_col='churn_label'
)

print(f"Transformed {X_quantum.shape[0]} samples with {X_quantum.shape[1]} features")

# ============================================================================
# QUANTUM PROCESSING: Execute quantum algorithms
# ============================================================================
dev = qml.device('default.qubit', wires=4)

@qml.qnode(dev)
def quantum_feature_extractor(params, x):
    """
    Quantum circuit for feature extraction from classical data.
    
    Args:
        params: Variational parameters for quantum feature map
        x: Classical data point (normalized to [0, 2*pi])
    
    Returns:
        array: Quantum features as expectation values
    """
    # Encode classical data into quantum state
    qml.AngleEmbedding(x, wires=range(4))
    
    # Apply variational quantum feature map
    qml.BasicEntanglerLayers(params, wires=range(4))
    
    # Extract quantum features
    return [qml.expval(qml.PauliZ(i)) for i in range(4)]

def process_batch_quantum(X_batch, params):
    """
    Process a batch of data through quantum circuit.
    
    Args:
        X_batch: Batch of classical data
        params: Quantum circuit parameters
    
    Returns:
        array: Quantum-processed features
    """
    quantum_features = []
    for x in X_batch:
        qf = quantum_feature_extractor(params, x)
        quantum_features.append(qf)
    return np.array(quantum_features)

# Initialize quantum parameters
quantum_params = np.random.uniform(0, 2*np.pi, (2, 4))

# Process data in batches (important for large datasets)
batch_size = 100
quantum_results = []
for i in range(0, len(X_quantum), batch_size):
    batch = X_quantum[i:i+batch_size]
    q_batch = process_batch_quantum(batch, quantum_params)
    quantum_results.append(q_batch)
    print(f"Processed batch {i//batch_size + 1}/{(len(X_quantum)-1)//batch_size + 1}")

quantum_features = np.vstack(quantum_results)
print(f"Quantum processing complete: {quantum_features.shape}")

# ============================================================================
# LOAD: Store results back to classical formats
# ============================================================================
def load_results(quantum_features, original_data, output_format='dataframe'):
    """
    Load quantum processing results back into classical data formats.
    
    Args:
        quantum_features: Quantum-processed features
        original_data: Original DataFrame for merging
        output_format: Output format ('dataframe', 'json', 'database')
    
    Returns:
        DataFrame or dict: Results in requested format
    """
    # Create results DataFrame
    q_cols = [f'quantum_feature_{i}' for i in range(quantum_features.shape[1])]
    results_df = pd.DataFrame(quantum_features, columns=q_cols)
    
    # Merge with original data
    results_df = pd.concat([original_data.reset_index(drop=True), 
                           results_df], axis=1)
    
    if output_format == 'dataframe':
        return results_df
    elif output_format == 'json':
        return results_df.to_dict('records')
    elif output_format == 'database':
        engine = create_engine('sqlite:///quantum_results.db')
        results_df.to_sql('quantum_features', engine, if_exists='replace')
        return results_df
    
    return results_df

# Load results
final_results = load_results(quantum_features, raw_data, output_format='dataframe')

# Save to CSV for downstream processing
final_results.to_csv('quantum_enhanced_features.csv', index=False)
print(f"Results saved: {len(final_results)} records with quantum features")

# ============================================================================
# INTEGRATION: Connect to downstream data engineering workflows
# ============================================================================
# Results can now be used in:
# - Data warehouses (BigQuery, Snowflake, Redshift)
# - Business intelligence tools (Tableau, Power BI)
# - Machine learning pipelines (scikit-learn, XGBoost)
# - Real-time analytics (Kafka, Spark Streaming)
# - Data lakes (S3, Azure Data Lake)

# Example: Feed into classical ML pipeline
from sklearn.ensemble import RandomForestClassifier
from sklearn.model_selection import train_test_split

# Use quantum features for classification
X_final = final_results[q_cols].values
y_final = final_results['churn_label'].values

X_train, X_test, y_train, y_test = train_test_split(
    X_final, y_final, test_size=0.2, random_state=42
)

clf = RandomForestClassifier(n_estimators=100, random_state=42)
clf.fit(X_train, y_train)
accuracy = clf.score(X_test, y_test)
print(f"Model accuracy with quantum features: {accuracy:.4f}")
\end{lstlisting}

This ETL pipeline example demonstrates how PennyLane enables quantum computing to be integrated into standard data engineering workflows used in business analytics and organizational data analysis. The pipeline follows the familiar Extract-Transform-Load pattern, making it accessible to data engineers and analysts who work with classical data processing tools. The quantum processing stage acts as an intermediate transformation that can enhance classical features with quantum properties, potentially providing advantages for certain machine learning tasks.

The pipeline's modular design enables integration with various data engineering pathways commonly used in organizational data analysis and business analytics applications. Quantum-processed results can be loaded into data warehouses (BigQuery, Snowflake, Redshift) for business intelligence applications, enabling analysts to query quantum-enhanced features using standard SQL interfaces. The results can be fed into real-time analytics systems (Apache Kafka, Spark Streaming) for operational decision-making, where quantum feature extraction can be performed on streaming data to enable real-time pattern recognition and anomaly detection. For large-scale analytics, quantum-processed data can be stored in data lakes (Amazon S3, Azure Data Lake) alongside classical data, enabling hybrid quantum-classical analytics workflows that leverage both types of features.

This ETL pattern also integrates with modern data orchestration tools such as Apache Airflow, Prefect, and Dagster, where quantum processing stages can be scheduled alongside classical transformations. The pipeline can be containerized using Docker and deployed on Kubernetes clusters, enabling scalable quantum data processing that matches the elasticity requirements of cloud-based data engineering infrastructure. For organizations using MLOps platforms (MLflow, Kubeflow, Weights \& Biases), quantum feature extraction can be tracked and versioned alongside classical preprocessing steps, ensuring reproducibility and enabling A/B testing of quantum-enhanced versus classical-only pipelines. This flexibility makes quantum-enhanced data processing compatible with existing organizational data infrastructure, enabling incremental adoption of quantum computing capabilities without requiring complete workflow redesign, while providing a clear path for scaling quantum data processing as hardware becomes more accessible and cost-effective.

\subsection{Integration with PyTorch}

PyTorch integration enables seamless construction of hybrid quantum-classical models with automatic differentiation:

Listing 11: PyTorch integration for hybrid models

\begin{lstlisting}[caption={Training quantum circuits with PyTorch}]
"""
Demonstrate integration of PennyLane with PyTorch for training
quantum circuits using PyTorch's optimization infrastructure.
"""
import pennylane as qml
import torch
import torch.nn as nn
from torch.utils.data import DataLoader, TensorDataset

# Quantum device with PyTorch interface
dev = qml.device('default.qubit', wires=2)

@qml.qnode(dev, interface='torch', diff_method='backprop')
def quantum_circuit(params, x):
    """
    Quantum circuit compatible with PyTorch automatic differentiation.
    
    Args:
        params: Trainable quantum parameters (PyTorch tensor)
        x: Input data (PyTorch tensor)
    
    Returns:
        tensor: Expectation value measurement
    """
    qml.RY(x[0], wires=0)
    qml.RY(x[1], wires=1)
    qml.RY(params[0], wires=0)
    qml.CNOT(wires=[0, 1])
    qml.RY(params[1], wires=1)
    return qml.expval(qml.PauliZ(0))

# Create dataset
X = torch.randn(100, 2)
y = (X.sum(dim=1) > 0).float()
dataset = TensorDataset(X, y)
loader = DataLoader(dataset, batch_size=10)

# Quantum parameters
quantum_params = nn.Parameter(torch.randn(2))

# Training loop
optimizer = torch.optim.Adam([quantum_params], lr=0.1)
criterion = nn.BCELoss()

for epoch in range(20):
    total_loss = 0
    for batch_x, batch_y in loader:
        optimizer.zero_grad()
        outputs = torch.stack([quantum_circuit(quantum_params, x) 
                              for x in batch_x])
        outputs = torch.sigmoid(outputs)
        loss = criterion(outputs, batch_y)
        loss.backward()
        optimizer.step()
        total_loss += loss.item()
    print(f"Epoch {epoch}: Loss = {total_loss/len(loader):.4f}")
\end{lstlisting}

This PyTorch integration example demonstrates how PennyLane enables quantum-enhanced deep learning for data science workflows and business analytics applications. The ability to use PyTorch's automatic differentiation with quantum circuits enables researchers to build end-to-end trainable models that combine quantum and classical components. The integration with PyTorch's data loading and optimization infrastructure demonstrates how PennyLane facilitates quantum machine learning within established deep learning workflows, making quantum computing accessible to researchers working in organizational data analysis and AI applications. This approach is useful for business analytics applications where hybrid quantum-classical models may provide advantages for specific learning tasks, enabling organizations to explore quantum-enhanced machine learning for predictive analytics, customer segmentation, or recommendation systems.

\subsection{Integration with TensorFlow}

TensorFlow integration provides similar capabilities for building hybrid models:

Listing 12: TensorFlow integration

\begin{lstlisting}[caption={Quantum circuits with TensorFlow}]
import pennylane as qml
import tensorflow as tf

# Quantum device with TensorFlow interface
dev = qml.device('default.qubit', wires=2)

@qml.qnode(dev, interface='tf', diff_method='backprop')
def quantum_layer(params, x):
    qml.RY(x[0], wires=0)
    qml.RY(x[1], wires=1)
    qml.RY(params[0], wires=0)
    qml.CNOT(wires=[0, 1])
    return qml.expval(qml.PauliZ(0))

# Create TensorFlow model
class QuantumModel(tf.keras.Model):
    def __init__(self):
        super().__init__()
        self.dense = tf.keras.layers.Dense(2)
        self.quantum_params = tf.Variable(tf.random.normal([1]))
        
    def call(self, inputs):
        x = self.dense(inputs)
        q_out = tf.stack([quantum_layer(self.quantum_params, x[i]) 
                         for i in range(tf.shape(x)[0])])
        return tf.nn.sigmoid(q_out)

# Training
model = QuantumModel()
model.compile(optimizer='adam', loss='binary_crossentropy')

# Dummy data
X = tf.random.normal((100, 4))
y = tf.cast(tf.reduce_sum(X, axis=1) > 0, tf.float32)

model.fit(X, y, epochs=10, verbose=1)
\end{lstlisting}

\subsection{Integration with JAX}

JAX integration enables high-performance quantum computing with just-in-time compilation:

Listing 13: JAX integration for high-performance quantum computing

\begin{lstlisting}[caption={Quantum circuits with JAX}]
import pennylane as qml
import jax
import jax.numpy as jnp
from jax import grad

# Quantum device with JAX interface
dev = qml.device('default.qubit', wires=2)

@qml.qnode(dev, interface='jax', diff_method='backprop')
def quantum_circuit(params, x):
    qml.RY(x[0], wires=0)
    qml.RY(x[1], wires=1)
    qml.RY(params[0], wires=0)
    qml.CNOT(wires=[0, 1])
    return qml.expval(qml.PauliZ(0))

# JAX-optimized cost function
def cost(params, x, target):
    return (quantum_circuit(params, x) - target)**2

# Compute gradient
cost_grad = grad(cost, argnums=0)

# Vectorized computation
vmap_circuit = jax.vmap(quantum_circuit, in_axes=(None, 0))

# Batch processing
params = jnp.array([0.5])
X = jnp.array([[0.1, 0.2], [0.3, 0.4], [0.5, 0.6]])
outputs = vmap_circuit(params, X)
print(f"Batch outputs: {outputs}")
\end{lstlisting}

\subsection{Integration with Scikit-learn}

PennyLane can be integrated with scikit-learn for quantum-enhanced machine learning:

Listing 14: Quantum-enhanced scikit-learn pipeline

\begin{lstlisting}[caption={Quantum feature engineering with scikit-learn}]
import pennylane as qml
from pennylane import numpy as np
from sklearn.pipeline import Pipeline
from sklearn.preprocessing import StandardScaler
from sklearn.ensemble import RandomForestClassifier
from sklearn.model_selection import train_test_split
from sklearn.datasets import make_classification

# Generate data
X, y = make_classification(n_samples=200, n_features=4, random_state=42)
X_train, X_test, y_train, y_test = train_test_split(X, y, test_size=0.3, 
                                                   random_state=42)

# Quantum feature map
dev = qml.device('default.qubit', wires=4)

@qml.qnode(dev)
def quantum_features(x):
    qml.AngleEmbedding(x, wires=range(4))
    qml.BasicEntanglerLayers(np.ones((1, 4)), wires=range(4))
    return [qml.expval(qml.PauliZ(i)) for i in range(4)]

# Create quantum feature transformer
class QuantumFeatureTransformer:
    def fit(self, X, y=None):
        return self
    
    def transform(self, X):
        return np.array([quantum_features(x) for x in X])

# Build pipeline
pipeline = Pipeline([
    ('scaler', StandardScaler()),
    ('quantum', QuantumFeatureTransformer()),
    ('classifier', RandomForestClassifier(n_estimators=100, random_state=42))
])

# Train and evaluate
pipeline.fit(X_train, y_train)
accuracy = pipeline.score(X_test, y_test)
print(f"Pipeline accuracy: {accuracy:.4f}")
\end{lstlisting}

\section{Discussion}

PennyLane provides a powerful tool for hybrid quantum-classical machine learning in Python-based research workflows, serving as a component for quantum-enhanced data science pipelines, business analytics applications, and AI dataset construction. This section examines the advantages and limitations of PennyLane, situating it within the broader context of quantum computing tools and identifying opportunities for future development.

\subsection{Advantages}

The primary advantage of PennyLane is its seamless integration between quantum computing and classical machine learning, making it a useful tool for data science workflows and business analytics applications. Unlike hardware-focused frameworks, PennyLane provides a device-agnostic interface that enables researchers to develop algorithms on simulators before deploying to quantum hardware. The automatic differentiation capabilities enable efficient gradient computation without manual derivation, reducing the complexity of variational quantum algorithm development.

Portability represents another advantage. PennyLane's device abstraction ensures that quantum circuits can be executed across different backends without code modification. This portability is important for reproducible research, where code must run consistently across diverse quantum computing environments. Unlike platform-specific solutions, PennyLane provides a standardized interface that works across simulators, cloud quantum services, and hardware backends.

Performance is also an advantage, particularly for gradient computation and optimization in quantum machine learning workflows. The support for multiple differentiation methods (parameter-shift, finite-difference, backpropagation) allows researchers to choose the most efficient approach for their specific use case. The integration with JAX enables just-in-time compilation and vectorization, providing performance benefits for large-scale quantum machine learning applications.

Integration capabilities enhance PennyLane's utility for data science workflows and business analytics applications. The library works seamlessly with popular machine learning frameworks such as PyTorch, TensorFlow, and JAX, enabling researchers to build hybrid quantum-classical models within established ML workflows. This integration reduces the need for custom quantum-classical interfaces and promotes the use of established libraries, ensuring reproducible research practices.

\subsection{Limitations}

Despite its advantages, PennyLane has several limitations that researchers should consider. The most significant limitation is the computational overhead of quantum circuit simulation, which can be constraining for large-scale applications. While PennyLane provides efficient simulators, the exponential scaling of quantum state space limits the size of circuits that can be efficiently simulated classically.

Hardware access represents another limitation. While PennyLane supports various hardware backends, access to quantum hardware may be limited or costly for some researchers. The queue times and noise characteristics of current quantum devices can make hardware execution impractical for many research workflows, limiting the ability to validate algorithms on real quantum systems. Error mitigation techniques~\cite{noise_mitigation} can help address noise issues, but they add computational overhead.

The learning curve for quantum computing concepts can also be a limitation. Researchers new to quantum computing must understand concepts such as quantum gates, measurements, and variational circuits before effectively using PennyLane. Additionally, researchers should be aware of challenges such as barren plateaus~\cite{barren_plateaus} in quantum neural network training, which can make optimization difficult for certain circuit architectures. While the library provides good documentation, the underlying quantum mechanics concepts require specialized knowledge that may not be familiar to all researchers working in data science or business analytics.

\subsection{Future Directions}

The future development of quantum machine learning tools is likely to focus on improved hardware integration, enhanced optimization techniques, and better support for large-scale applications. As quantum hardware improves, extending PennyLane's capabilities to leverage new hardware features becomes important for quantum-enhanced data science and business analytics applications.

Improved optimization algorithms represent another area for future development. Current classical optimizers may not be optimal for the specific landscape of quantum cost functions. Developing quantum-aware optimization techniques could improve convergence and solution quality for variational quantum algorithms.

Integration with cloud computing and distributed systems could also enhance PennyLane's capabilities for large-scale quantum machine learning applications. As research datasets grow, the ability to distribute quantum circuit evaluation across multiple devices or cloud resources becomes important for practical applications.

\section{Conclusion}

PennyLane represents a powerful component for hybrid quantum-classical machine learning, providing quantum circuit construction capabilities that enable portable and reproducible workflows for data science pipelines, business analytics applications, and AI dataset construction projects. Through this examination, we have demonstrated PennyLane's versatility across diverse research domains, from quantum machine learning and optimization to quantum chemistry and finance, establishing it as a tool for quantum-enhanced computational research workflows.

The library's integration with classical ML frameworks makes it a useful tool for researchers working at the intersection of quantum computing and machine learning, while its device abstraction and automatic differentiation capabilities enable the creation of analytical pipelines. The portability and cross-platform compatibility of PennyLane contribute to reproducible research practices, ensuring that code can be shared and executed consistently across different quantum computing environments.

Despite its limitations, including computational overhead and hardware access constraints, PennyLane provides sufficient functionality for the majority of hybrid quantum-classical machine learning needs in computational research. The library's performance characteristics and integration capabilities make it suitable for processing quantum-enhanced machine learning workflows, while its syntax reduces the learning curve for researchers new to quantum computing.

\subsection{Practical Recommendations}

For researchers incorporating PennyLane into their workflows, several practical recommendations emerge. First, researchers should start with simulators before moving to hardware, as this allows for algorithm development and debugging without the constraints of quantum hardware. The \texttt{default.qubit} device provides a fast, reliable simulator for initial development.

Second, researchers should leverage PennyLane's integration with classical ML frameworks when building hybrid models. The ability to combine quantum and classical layers within established ML workflows enables researchers to incrementally explore quantum enhancements without completely rewriting existing code.

Third, researchers should consider the choice of differentiation method based on their specific use case. Parameter-shift rules work well for hardware execution, while backpropagation provides efficiency advantages for simulators. Understanding these trade-offs improves the efficiency of quantum machine learning workflows.

Fourth, researchers should document their quantum circuit architectures and hyperparameters to improve reproducibility. Including circuit diagrams, parameter initialization strategies, and optimization settings in research documentation enables other researchers to understand and extend existing quantum machine learning workflows.

\subsection{Future Directions}

The future of hybrid quantum-classical machine learning is likely to involve improved hardware integration, enhanced optimization techniques, and better support for large-scale applications. As quantum hardware matures and becomes more accessible, the ability to deploy PennyLane-based algorithms on real quantum devices will become increasingly important.

The development of quantum-aware optimization techniques and improved classical-quantum interfaces represents an opportunity for the research community. Such developments would enable researchers to more effectively leverage quantum computing for machine learning applications, potentially providing advantages for specific problem domains.

Additionally, the integration of quantum machine learning with cloud computing and distributed systems could enable more scalable quantum-enhanced analytics. This could be useful for research involving large datasets where quantum feature maps or quantum kernels may provide advantages over classical methods.

In conclusion, PennyLane serves as a powerful component that enables reproducible and accessible hybrid quantum-classical machine learning in computational research. By providing a simple interface for quantum circuit construction and integration with classical ML frameworks, PennyLane contributes to the broader goal of making quantum-enhanced machine learning more accessible, reproducible, and efficient. As the research community continues to develop new quantum algorithms and hardware, PennyLane will remain a component of the Python ecosystem for quantum machine learning and hybrid quantum-classical workflows.


\begin{thebibliography}{25}
\bibitem{pennylane}
Bergholm, V., Izaac, J., Schuld, M., Gogolin, C., Ahmed, S., Ajith, V., \ldots{} Killoran, N. (2022). PennyLane: Automatic differentiation of hybrid quantum-classical computations. arXiv preprint arXiv:1811.04968.

\bibitem{pennylane_recent}
PennyLane AI. (2024). PennyLane Documentation. \url{https://docs.pennylane.ai/}

\bibitem{qiskit}
ANIS, M. S., et al. (2021). Qiskit: An Open-source Framework for Quantum Computing. Qiskit/qiskit: Qiskit 0.45.0. Zenodo. \url{https://doi.org/10.5281/zenodo.2573505}

\bibitem{qiskit_ml}
Lorenz, R., Pearson, A., Meichanetzidis, K., Kotecha, K., \& Meyer, H. M. (2021). Qiskit Machine Learning: A quantum machine learning framework. arXiv preprint arXiv:2111.08467.

\bibitem{cirq}
Cirq Developers. (2021). Cirq: A Python framework for creating, editing, and invoking Noisy Intermediate Scale Quantum (NISQ) circuits. \url{https://github.com/quantumlib/Cirq}

\bibitem{tfq}
Broughton, M., Verdon, G., McCourt, T., Martinez, A. J., Yoo, J. H., Isakov, S. V., \ldots{} Babbush, R. (2020). TensorFlow Quantum: A software framework for quantum machine learning. arXiv preprint arXiv:2003.02989.

\bibitem{quantum_kernel}
Havlíček, V., Córcoles, A. D., Temme, K., Harrow, A. W., Kandala, A., Chow, J. M., \& Gambetta, J. M. (2019). Supervised learning with quantum-enhanced feature spaces. Nature, 567(7747), 209-212.

\bibitem{quantum_kernel_recent}
Schuld, M., \& Killoran, N. (2022). Is quantum advantage the right goal for quantum machine learning? PRX Quantum, 3(3), 030101.

\bibitem{quantum_ml_survey}
Biamonte, J., Wittek, P., Pancotti, N., Rebentrost, P., Wiebe, N., \& Lloyd, S. (2017). Quantum machine learning. Nature, 549(7671), 195-202.

\bibitem{quantum_ml_review}
Cerezo, M., Arrasmith, A., Babbush, R., Benjamin, S. C., Endo, S., Fujii, K., \ldots{} Coles, P. J. (2021). Variational quantum algorithms. Nature Reviews Physics, 3(9), 625-644.

\bibitem{vqe}
Peruzzo, A., McClean, J., Shadbolt, P., Yung, M. H., Zhou, X. Q., Love, P. J., \ldots{} O'Brien, J. L. (2014). A variational eigenvalue solver on a photonic quantum processor. Nature Communications, 5(1), 4213.

\bibitem{vqe_recent}
Tilly, J., Chen, H., Cao, S., Picozzi, D., Setia, K., Li, Y., \ldots{} Grant, E. (2022). The variational quantum eigensolver: A review of methods and best practices. Physics Reports, 986, 1-128.

\bibitem{hybrid_ml}
Benedetti, M., Lloyd, E., Sack, S., \& Fiorentini, M. (2019). Parameterized quantum circuits as machine learning models. Quantum Science and Technology, 4(4), 043001.

\bibitem{qaoa}
Farhi, E., Goldstone, J., \& Gutmann, S. (2014). A quantum approximate optimization algorithm. arXiv preprint arXiv:1411.4028.

\bibitem{qaoa_recent}
Akshay, V., Philathong, H., Morales, M. E., \& Biamonte, J. (2020). Reachability deficits in quantum approximate optimization. Physical Review Letters, 124(9), 090504.

\bibitem{data_encoding}
Schuld, M., Bocharov, A., Svore, K. M., \& Wiebe, N. (2020). Circuit-centric quantum classifiers. Physical Review A, 101(3), 032308.

\bibitem{portfolio_quantum}
Rebentrost, P., \& Lloyd, S. (2018). Quantum computational finance: quantum algorithm for portfolio optimization. arXiv preprint arXiv:1811.03975.

\bibitem{portfolio_recent}
Herman, D., Googin, C., Liu, X., Galda, A., Safro, I., Sun, Y., \ldots{} Pistoia, M. (2022). A survey of quantum computing for finance. arXiv preprint arXiv:2201.02773.

\bibitem{gradient_methods}
Mari, A., Bromley, T. R., Izaac, J., Schuld, M., \& Killoran, N. (2021). Transfer learning in hybrid classical-quantum neural networks. Quantum, 4, 340.

\bibitem{barren_plateaus}
McClean, J. R., Boixo, S., Smelyanskiy, V. N., Babbush, R., \& Neven, H. (2018). Barren plateaus in quantum neural network training landscapes. Nature Communications, 9(1), 4812.

\bibitem{expressibility}
Sim, S., Johnson, P. D., \& Aspuru-Guzik, A. (2019). Expressibility and entangling capability of parameterized quantum circuits for hybrid quantum-classical algorithms. Advanced Quantum Technologies, 2(12), 1900070.

\bibitem{noise_mitigation}
Temme, K., Bravyi, S., \& Gambetta, J. M. (2017). Error mitigation for short-depth quantum circuits. Physical Review Letters, 119(18), 180509.

\bibitem{quantum_advantage}
Huang, H. Y., Broughton, M., Mohseni, M., Babbush, R., Boixo, S., Neven, H., \& McClean, J. R. (2021). Power of data in quantum machine learning. Nature Communications, 12(1), 2631.

\bibitem{nisk_era}
Preskill, J. (2018). Quantum computing in the NISQ era and beyond. Quantum, 2, 79.

\bibitem{quantum_chemistry_recent}
McArdle, S., Endo, S., Aspuru-Guzik, A., Benjamin, S. C., \& Yuan, X. (2020). Quantum computational chemistry. Reviews of Modern Physics, 92(1), 015003.
\end{thebibliography}
\end{document}